\documentclass[a4paper]{article}

\begin {document}

\title {\bf The Quantum Reduced Action In Higher Dimensions}
\author{A.~Bouda\footnote{Electronic address:
{\tt bouda\_a@yahoo.fr}}
\\
Laboratoire de Physique Th\'eorique, Universit\'e de B\'eja\"\i a,\\
Route Targa Ouazemour, 06000 B\'eja\"\i a, Algeria\\}

\date{\today}

\maketitle

\begin{abstract}
\noindent
The solution with respect to the reduced action of the one-dimensional stationary
quantum Hamilton-Jacobi equation is well known in the literature. The extension to
higher dimensions in the separated variable case was proposed in contradictory
formulations. In this paper we provide new insights into the construction of the
reduced action. In particular, contrary to the classical mechanics case, we analytically
show that the reduced action constructed as a sum of one variable functions does not
contain a complete information about the quantum motion. In the same context,
we also make some observations about recent results concerning quantum trajectories.
Finally, we will examine the conditions in which microstates appear even in the
case where the wave function is complex.
\end{abstract}

\vskip\baselineskip

\noindent
PACS: 03.65.Ca; 03.65.Ta; 02.30.Jr

\noindent
Key words:  reduced action, quantum stationary Hamilton-Jacobi equation,
separated variables, trajectories, microstates.

\newpage

\section{Introduction}

In the context of quantum trajectories, the notion of the action was first introduced by
de Broglie \cite{broglie} and Bohm \cite{bohm1,bohm2} by writing the wave function in the well-known
form
\begin{equation}
\Psi =R \ \exp\left(i{{S}\over{\hbar}}\right) \ ,
\end{equation}
where $R$ and $S$ are real functions. The substitution of this expression in the
time-dependent Schr\"odinger equation leads to the two following partial derivative
equations
\begin {equation}
 {1\over 2m}\left(\vec\nabla S\right)^2-
{\hbar^2\over 2m}{\Delta R\over R}+V = -{\partial S\over \partial t} \ ,
\end {equation}
\begin {equation}
{\partial R^2\over \partial t} + \vec {\nabla}\cdot(R^2\vec{\nabla}S)=0 \ ,
\end {equation}
$V$ being an external potential. Relation (3) is the continuity equation. Relation (2)
reminds us of the Hamilton-Jacobi equation with an additional term
\begin {equation}
U=-{\hbar^2\over 2m}{\Delta R\over R}\ ,
\end {equation}
called the quantum potential and to which it is attributed the quantum effects. The function
$S$ is  identified as the quantum action and relation (2) is then called the quantum
Hamilton-Jacobi equation (QHJE). In the stationary case, we have
\begin{equation}
S(x,y,z,t)=S_0(x,y,z)-Et \ ,
\end{equation}
and
\begin{equation}
\Psi(x,y,z,t) =\exp\left(-{i\over{\hbar}}Et \right)\phi(x,y,z) \ ,
\end{equation}
$E$ being the energy and $S_0$ the quantum reduced action. By considering a system
described by a wave function such that $\phi(x,y,z)$ is real up to a constant phase factor,
relations (1), (5) and (6) show clearly that $S_0$ is constant. This feature is unsatisfactory.
In the context of the equivalence postulate of quantum mechanics \cite{FM1a,FM1b,FM2,BFM} from
which the Schr\"odinger equation (SE) was reproduced, this difficulty was surmounted
by showing that the wave function takes the form
\begin{equation}
\phi(x,y,z) =R\left[\alpha \ \exp\left(i{{S_0}\over{\hbar}}\right)
+\beta \ \exp\left(-i{{S_0}\over{\hbar}}\right)\right] \ ,
\end{equation}
where $\alpha$ and $\beta$ are complex constants. This bipolar form was also used by
Poirier \cite{poirier} in order to reconcile semiclassical and Bohmian Mechanics.
The new functions $R$ and $S_0$ satisfy the equations
\begin {equation}
 {1\over 2m}\left(\vec\nabla S_{0}\right)^2-
{\hbar^2\over 2m}{\Delta R\over R}+V(x,y,z)=E \ ,
\end {equation}
\begin {equation}
\vec {\nabla}\cdot(R^2\vec{\nabla}S_{0})=0 \ ,
\end {equation}
which are fundamentally different from the stationary version of (2) and (3) since
in the two cases the couple $(R, S_0)$ is linked in different manners to the wave
function. To perceive this difference, one can realize the form (7) guarantees
that $S_0$ is never constant even in the case where the wave function is real, up to
a constant phase factor. These results were reproduced in \cite{bouda1} by
appealing to the probability current.

In one dimension, Eqs. (8) and (9) turn out to be
\begin{eqnarray}
{1\over2m}{\left(\partial{S_0}\over\partial{x}\right)^2}+
V\left(x\right)-E=\hskip60mm&& \nonumber \\
{\hbar ^2\over4m}\left[{{3\over2}
\left(\partial{S_0}\over{\partial{x}}\right)^{-2}
\left(\partial^2{S_0}\over\partial{x^2}\right)^2}-
{\left(\partial{S_0}\over\partial{x}\right)^{-1}}
{\left(\partial^3{S_0}\over\partial{x^3}\right)}\right] \ .
\end{eqnarray}
The solution of this equation for an arbitrary potential $V(x)$ has been investigated
in \cite{floyd1a,floyd1b} and it is written as
\begin{equation}
S_0=\hbar \hskip2pt \arctan\left({b (\phi_1 / \phi_2) +c/2
\over (ab-c^2/4)^{1/2} } \right)+ K \ ,
\end{equation}
where $(a,b,c,K)$ is a set of real constants such that $a>0$, $b>0$ and $ab>c^2/4$.
The couple $(\phi_1,\phi_2)$ is a set of real independent solutions of the associated SE.
The above expression of $S_0$ solves (10) if the Wronskian $W=\phi_1\phi_2'-\phi_1'\phi_2$
is scaled so that $W^2 \equiv 2m/[\hbar^2(ab -c^2/4)]$. As (10) is a second order differential
equation with respect to $\partial{S_0}/\partial{x}$, the expression of $S_0$ must contain
two integration constants on top of the additive one. This is the reason for which the
three parameters $(a,b,c)$ are not independent since they are linked to $W$ for a
given choice of the couple $(\phi_1,\phi_2)$. Therefore, it is possible to eliminate one of them.
That's what is done in \cite{bouda1} where the solution of (10) is written as
\begin{equation}
S_0=\hbar \hskip2pt \arctan\left({\phi_1+\nu \phi_2}
\over{\mu \phi_1+\phi_2}\right)+\hbar l \ ,
\end{equation}
$(\mu,\nu,l)$ are independent parameters playing the role of integration constants.
They must satisfy the condition $\mu \nu \neq 1$ in order to guarantee that $S_0$ never takes a
constant value. Expression (12) solves (10) without any condition.
It is shown in \cite{BD1} that solution (12) can be written in the form
\begin{equation}
S_0=\hbar \hskip2pt \arctan\left(\mu {\phi_1\over \phi_2} + \nu \right)+ \hbar l \ ,
\end{equation}
with suitable redefining of $\mu$, $\nu$, $\phi_1$ and $\phi_2$. Another equivalent expression
for $S_0$ was proposed in \cite{FM2}.

In \cite{BM1} and \cite{djama1}, the extension of this solution to higher dimensions in
the separated variable case is investigated and the results are contradictory. The aim of this
paper is to provide new insights into the construction of the reduced action in higher dimensions.
In section 2, we discuss the link between the reduced action and the wave function.
In section 3, we will provide several arguments to show that the usual method which consists
in assuming the reduced action as a sum of one variable functions, as in classical mechanics,
leads to an incomplete solution. In section 4, we make some remarks about quantum
trajectories and microstates. Section 5 is devoted to conclusion.

\section{ The reduced action and the wave function}

Let us consider for stationary states the separated variable case where the potential takes
the following form
\begin{equation}
V(x,y,z) = V_{x}(x) + V_{y}(y) + V_{z}(z) \ .
\end{equation}
Writing the solution for the SE,
\begin{equation}
-{\hbar^{2} \over 2m} \Delta \phi(x,y,z) + V(x,y,z) \phi(x,y,z)=E \phi(x,y,z) \ ,
\end{equation}
in the form
\begin{equation}
\phi (x,y,z) = \phi_x(x)\phi_y(y)\phi_z(z) \ ,
\end{equation}
we deduce that
\begin{equation}
-{\hbar^{2} \over 2m}{d^{2} \phi_x \over {dx^{2}}}+V_{x} \phi_x=E_{x} \phi_x \ ,
\end{equation}
\begin{equation}
-{ \hbar^{2} \over 2m}{d^{2} \phi_y \over {dy^{2}}}+V_{y} \phi_y = E_{y} \phi_y \ ,
\end{equation}
\begin{equation}
-{\hbar^{2} \over 2m}{d^{2} \phi_z \over {dz^{2}}}+V_{z} \phi_z = E_{z} \phi_z \ ,
\end{equation}
where $E_{x}$, $E_{y}$ and $E_{z}$ are real constants satisfying
\begin{equation}
E_{x}+E_{y}+E_{z}=E \ .
\end{equation}
In this section, we will make some comments about ref. \cite{djama1}.

The first comment is just of a pedagogical nature. It concerns the
establishment of the above relations (17), (18) and (19). In section 2 of ref. \cite{djama1},
the author considered the above relation (20) as an hypothesis and substituted it in (15)
in order to obtain (17), (18) and (19). This manner is wrong because $E_{x}$, $E_{y}$
and $E_{z}$ follow from the procedure of variable separation and as well-known they are
integration constants.

Another comment concerns the form of the three functions $\phi_q(q)$ which are written in
section 2 of ref. \cite{djama1} as
\begin{equation}
\phi_q(q) =R_q(q)\left[\alpha_q \ \exp\left(i{{S_{0q}(q)}\over{\hbar}}\right)
+\beta_q \ \exp\left(-i{{S_{0q}(q)}\over{\hbar}}\right)\right] \ ,
\end{equation}
$R_q(q)$ and $S_{0q}(q)$ being real functions, $\alpha_q $ and $\beta_q $ complex
constants and $q=x,y,z$. Substituting (21) in (16), we obtain
\begin{eqnarray}
\phi(x,y,z) = R_x R_y R_z  \left\{ \alpha_x\alpha_y\alpha_z  \
     \exp\left[{i\over{\hbar}}\left(S_{0x}+S_{0y}+S_{0z}\right)\right]  \right. \hskip20mm&& \nonumber\\
                            \left. + \alpha_x\alpha_y \beta_z \
      \exp\left[{i\over{\hbar}}\left(S_{0x}+S_{0y}-S_{0z}\right)\right] + \cdots \right\} \ ,
\end{eqnarray}
where the six missing terms can be easily completed. Whatever the expression of the
reduced action $S_0$ in terms of $S_{0x}$, $S_{0y}$ and $S_{0z}$, we see that the form
(22) cannot reproduce (7).

In addition, it is stated in \cite{djama1} that the above form (21) is
justified in \cite{FM1a,FM1b}. We emphasize that this is wrong. In \cite{FM1a,FM1b}, the form (21) is
established in the context of the equivalence postulate of Faraggi and Matone for the
one-dimensional case. In higher dimensions, the relation between the wave
function and the couple $(R,S_0)$ is established in the same context in \cite{BFM}
and it is given by the above Eq. (7). Admittedly the separated variable case is particular
but relation (7) continues to work and never reduces to (22) which follows
straightforwardly from (21). Thus, contrary to the statement made in \cite{djama1},
relations (21) and (22) are not in agreement with the equivalence postulate of
quantum mechanics \cite{FM1a,FM1b,FM2,BFM} as they are not with Bohmian mechanics.

The last comment is to say that the procedure used in the construction of the reduced
action in \cite{djama1} leads to a result which does not contain a complete information
about the quantum motion. To justify this, some mathematical details are necessary.
We come back to this feature in the next section.

We would like to add that all these weakness are reproduced in the others
sections of \cite{djama1} where spherical and cylindrical symmetries are considered
and in \cite{djama2} devoted to the hydrogen atom.

\section {The reduced action as a solution of the QHJE}

Let us call
    $ \left( X_{1},X_{2} \right) $,
    $ \left( Y_{1},Y_{2} \right) $
and
    $ \left( Z_{1},Z_{2} \right) $
three couples of real independent solutions respectively of
(17), (18) and (19). It follows that the three-dimensional SE
admits eight real independent solutions
\begin{eqnarray}
\left\{ \begin{array}{cc}
               \phi_{1}=X_{1}Y_{1}Z_{1}, \hskip6pt
               \phi_{2}=X_{1}Y_{1}Z_{2}, \hskip6pt
               \phi_{3}=X_{1}Y_{2}Z_{1}, \hskip6pt
               \phi_{4}=X_{1}Y_{2}Z_{2},    \\
               \phi_{5}=X_{2}Y_{1}Z_{1}, \hskip6pt
               \phi_{6}=X_{2}Y_{1}Z_{2}, \hskip6pt
               \phi_{7}=X_{2}Y_{2}Z_{1}, \hskip6pt
               \phi_{8}=X_{2}Y_{2}Z_{2}.
           \end{array}  \right.
\end{eqnarray}

By imposing the invariance of the reduced action, up to an additive constant, under any
linear transformation of the solutions of the SE, we showed from the above
relations (8) and (9) that the reduced action in the separated variable case
is given by \cite{BM1}
\begin{equation}
S_0^{(1)}= \hbar \hskip 2pt \arctan \left( {  {\sum_{i=1}^{8} \nu_i \phi_i}
 \over {\sum_{i=1}^{8}{\mu}_i\phi_i} } \right) + \hbar l \ ,
\end{equation}
in which we can fix freely one parameter among ${\nu_i}$ and one among ${\mu_i}$. The fourteen
remaining pertinent parameters play the role of integration constants.

In section 2 of ref. \cite{djama1}, the author wrote the reduced action in the following
form
\begin{equation}
S_0^{(2)}(x,y,z) = S_{0x}(x) + S_{0y}(y)  + S_{0z}(z) \ ,
\end{equation}
as in classical mechanics. In addition, it is assumed in \cite{djama1} that in this sum $S_{0x}$
is the solution of the usual one-dimensional QHJE, Eq. (10), as it is assumed for
$S_{0y}$ and $S_{0z}$ with analogous equations.
Let us show that the above expression (25) of $S_0^{(2)}(x,y,z)$ does not contain a
complete information about the motion of the particle and it is a particular
case of (24). For this purpose, by taking into account (12), let us write (25) as
\begin{eqnarray}
S_0^{(2)}=\hbar \arctan
    \left( { X_1+ \gamma_1 X_2\over \gamma_2 X_1 + X_2} \right)
    + \hbar \arctan
    \left({ Y_1 +\gamma_3 Y_2 \over \gamma_4 Y_1 +  Y_2  } \right) \hskip20mm \nonumber \\
   + \hbar \arctan
      \left({Z_1 + \gamma_5 Z_2
              \over \gamma_6 Z_1 +  Z_2 } \right) +\hbar l \ ,
\end{eqnarray}
in which $\gamma_1,...,\gamma_6$ are integration constants and $\hbar l$ represents the sum
of the three additive constants associated to $S_{0x}$, $S_{0y}$ and $S_{0z}$. Let us search
for a function $h$ defined by
\begin{equation}
 \arctan h = \arctan f_x + \arctan f_y + \arctan f_ z ,
\end{equation}
where
\begin{equation}
 f_x= { X_1+ \gamma_1 X_2\over \gamma_2 X_1 + X_2} \ ,
\end{equation}
\begin{equation}
f_y= { Y_1 +\gamma_3 Y_2 \over \gamma_4 Y_1 +  Y_2  } \ ,
\end{equation}
\begin{equation}
f_z= {Z_1 + \gamma_5 Z_2 \over \gamma_6 Z_1 +  Z_2 } \ .
\end{equation}
Knowing that
$
\tan (a+b)=(\tan a + \tan b)/(1- \tan a \tan b) ,
$
from (27) we deduce that
\begin{eqnarray}
h & = &
   \tan \left[ \arctan f_x + \arctan f_y + \arctan f_z  \right] \nonumber \\
  & =  & { {\tan (\arctan f_x + \arctan f_y) +f_ z}
       \over{1- \tan (\arctan f_x + \arctan f_y) f_ z }} \nonumber \\
  & = & {  (f_x +f_y) (1-f_x f_y)^{-1} +f_z
        \over 1-  (f_x +f_y )(1-f_x f_y )^{-1}f_z } \nonumber \\
  & = & {f_x +f_y+f_z-f_x f_y f_z \over 1-f_x f_y- f_x f_z -f_y f_z} \ .
\end{eqnarray}
Substituting $f_x$, $f_y$ and $f_z$ by their expressions (28),
(29) and (30), we obtain
\begin{eqnarray}
 h =
 \left[ { \over } ( X_1 + \gamma_1 X_2)
               (\gamma_4 Y_1 +  Y_2)
     (\gamma_6 Z_1 + Z_2)+   \right. \hskip20mm \nonumber \\
(\gamma_2 X_1 + X_2)(Y_1 + \gamma_3 Y_2)
     (\gamma_6 Z_1 + Z_2)+ \hskip20mm \nonumber \\
 (\gamma_2 X_1 + X_2)(\gamma_4 Y_1 + Y_2)
     (Z_1 + \gamma_5 Z_2)- \hskip20mm \nonumber \\
 \left. ( X_1 + \gamma_1 X_2)( Y_1 + \gamma_3 Y_2)
     ( Z_1 + \gamma_5 Z_2) { \over } \right] \times \hskip17mm \nonumber \\
\hskip25mm \left[ { \over }
   (\gamma_2 X_1 + X_2) (\gamma_4 Y_1 + Y_2)
     (\gamma_6 Z_1 + Z_2) -   \right. \nonumber \\
\hskip25mm  ( X_1 + \gamma_1 X_2)( Y_1 + \gamma_3 Y_2)
     (\gamma_6 Z_1 +  Z_2)- \nonumber \\
\hskip25mm  ( X_1 + \gamma_1 X_2)(\gamma_4 Y_1 +  Y_2)
     ( Z_1 + \gamma_5 Z_2)- \nonumber \\
 \left. (\gamma_2 X_1 +  X_2)( Y_1 + \gamma_3 Y_2)
     ( Z_1 + \gamma_5 Z_2)  { \over } \right]^{-1} \ , \hskip-3mm
\end{eqnarray}
which can be written in the form
\begin{equation}
h={{\sum_{i=1}^{8} \lambda_i \phi_i}
 \over {\sum_{i=1}^{8}{\delta}_i\phi_i} } \ ,
\end{equation}
where $\phi_i \ (i=1,...,8)$ are defined in (23). From (32),
the sixteen coefficients $\lambda_i$ and $\delta_i$  can be easily
expressed in terms of the six parameters $\gamma_1,...,\gamma_6$.
Taking into account relations (27), (28), (29), (30) and (33), expression
(26) turns out to be
\begin{equation}
S_0^{(2)}=\hbar \arctan
      \left({{\sum_{i=1}^{8} \lambda_i \phi_i}
 \over {\sum_{i=1}^{8}{\delta}_i\phi_i} } \right) +\hbar l \ ,
\end{equation}
Comparing (24) and (34), we see clearly that $S_0^{(1)}$ and $S_0^{(2)}$ have the
same form. However, in expression (24) there are fourteen independent parameters
among $(\nu_1,...,\nu_8)$ and $ (\mu_1,...,\mu_8)$ while in (34) there are six
independent parameters $\gamma_1,...,\gamma_6$. Thus, solution (34) can be
obtained by choosing particular values for eight $(8=14-6)$ parameters in (24).
This is the proof that the solution (25) is a particular case of (24).

Furthermore, we can show that it is (34) which is lacking in parameters and
not in (24) that there is a surplus. In fact, even by supposing that (25) is true,
we will show that Eqs. (8) and (9) lead to a solution more general than (26).
For this purpose, observe that expression (16) for the wave function
indicates that
\begin{equation}
R(x,y,z) = R_{x}(x)  R_{y}(y)  R_{z}(z) \ .
\end{equation}
This last expression is suggested in \cite{BFM} and implicitly assumed in \cite{djama1}.
So, by setting $S_0(x,y,z)$ as in (25) and $R(x,y,z)$ as in (35), we obtain from (8)
\begin{eqnarray}
{1\over 2m}\left({\partial S_{0x}\over \partial x}\right)^2+
{1\over 2m}\left({\partial S_{0y}\over \partial y}\right)^2+
{1\over 2m}\left({\partial S_{0z}\over \partial z}\right)^2
-{\hbar^2\over 2mR_x}{\partial^2 R_x\over \partial x^2} \hskip12mm \nonumber \\
-{\hbar^2\over 2mR_y}{\partial^2 R_y\over \partial y^2}
-{\hbar^2\over 2mR_z}{\partial^2 R_z\over \partial z^2}
+V_{x}(x) + V_{y}(y)+ V_{z}(z)=E \ ,
\end{eqnarray}
where we have used (14). The procedure of variable separation leads to
the three following equations
\begin{equation}
{1\over 2m}\left({\partial S_{0x}\over \partial x}\right)^2
-{\hbar^2\over 2mR_x}{\partial^2 R_x\over \partial x^2}
+V_{x}(x) =E_x,
\end{equation}
\begin{equation}
{1\over 2m}\left({\partial S_{0y}\over \partial y}\right)^2
-{\hbar^2\over 2mR_y}{\partial^2 R_y\over \partial y^2}
+ V_{y}(y)=E_y \ ,
\end{equation}
\begin{equation}
{1\over 2m}\left({\partial S_{0z}\over \partial z}\right)^2
-{\hbar^2\over 2mR_z}{\partial^2 R_z\over \partial z^2}
+ V_{z}(z)=E_z  \ .
\end{equation}
The three integration constants $E_x$, $E_y$ and $E_z$ satisfy
the condition (20). Substituting expressions (25) and (35) in Eq. (9)
and dividing then the obtained relation by $R^2$, we find
\begin{equation}
{1 \over R_x^2 }{\partial \over \partial x}
\left(R_x^2 {\partial S_{0x} \over \partial x }\right) +
{1 \over R_y^2 }{\partial \over \partial y}
\left(R_y^2 {\partial S_{0y} \over \partial y }\right) +
{1 \over R_z^2 }{\partial \over \partial z}
\left(R_x^2 {\partial S_{0z} \over \partial z }\right)
= 0 \ .
\end{equation}
From this relation, the procedure of variable separation leads to
\begin{equation}
{1 \over R_x^2 }{\partial \over \partial x}
\left(R_x^2 {\partial S_{0x} \over \partial x }\right) =c_1 \ ,
\end{equation}
\begin{equation}
{1 \over R_y^2 }{\partial \over \partial y}
\left(R_y^2 {\partial S_{0y} \over \partial y }\right) = c_2 \ ,
\end{equation}
\begin{equation}
{1 \over R_z^2 }{\partial \over \partial z}
\left(R_x^2 {\partial S_{0z} \over \partial z }\right) = c_3 \ .
\end{equation}
Taking into account Eq. (40), the integration constants
$c_1$, $c_2$ and  $c_3$ must satisfy the condition
\begin{equation}
c_1+ c_2+c_3=0  \ .
\end{equation}
With the use of $x$ as variable, the usual one-dimensional
case can be obtained from (41) by setting $c_1=0$.
The presence of the constants $c_1$, $c_2$ and  $c_3$ in Eqs. (41), (42) and (43) is the
reason for which the extension to three dimensions is not similar to classical mechanics.
In fact, rewriting (41) in the form
\begin{equation}
{2 \over R_x }{\partial R_x \over \partial x}
{\partial S_{0x} \over \partial x } +
{\partial^2 S_{0x} \over \partial x^2 } =c_1 \ ,
\end{equation}
and taking the derivative with respect to $x$, we find
\begin{equation}
{2 \over R_x }{\partial^2 R_x \over \partial x^2}
{\partial S_{0x} \over \partial x } -
{2 \over R^2_x }\left({\partial R_x \over \partial x}\right)^2
{\partial S_{0x} \over \partial x }+
{2 \over R_x }{\partial R_x \over \partial x}
{\partial^2 S_{0x} \over \partial x^2 }+
{\partial^3 S_{0x} \over \partial x^3 } =0 \ ,
\end{equation}
From (45), we have
\begin{equation}
{\partial R_x \over \partial x}=-{R_x \over 2 }
\left( {\partial S_{0x} \over \partial x }\right)^{-1}
\left({\partial^2 S_{0x} \over \partial x^2 } -c_1 \right) \ ,
\end{equation}
Substituting this expression in (46), we find
\begin{eqnarray}
{1 \over R_x }{\partial^2 R_x \over \partial x^2}=
-{1 \over 2 }\left( {\partial S_{0x} \over \partial x }\right)^{-1}
{\partial^3 S_{0x} \over \partial x^3 } + \hskip40mm \nonumber \\
{1 \over 4 } \left( {\partial S_{0x} \over \partial x }\right)^{-2}
\left[3\left({\partial^2 S_{0x} \over \partial x^2 }\right)^2 -
4c_1 {\partial^2 S_{0x} \over \partial x^2 } + c_1^2\right]  \ .
\end{eqnarray}
Using this result in (37), we obtain
\begin{eqnarray}
{1\over 2m} \left({\partial S_{0x}\over \partial x}\right)^2
-{\hbar^2\over 4m}
  \left[ {3 \over 2 } \left( {\partial S_{0x} \over \partial x }\right)^{-2}
\left({\partial^2 S_{0x} \over \partial x^2 }\right)^2
 -\left( {\partial S_{0x} \over \partial x }\right)^{-1}
{\partial^3 S_{0x} \over \partial x^3 } \right] \hskip8mm \nonumber \\
+V_{x}(x) -E_x ={\hbar^2 c_1\over 8m}
\left( {\partial S_{0x} \over \partial x }\right)^{-2}\left[
c_1-4 {\partial^2 S_{0x} \over \partial x^2 }  \right] \ .
\end{eqnarray}
In the same manner, we can also obtain
\begin{eqnarray}
{1\over 2m} \left({\partial S_{0y}\over \partial y}\right)^2
-{\hbar^2\over 4m}
  \left[ {3 \over 2 } \left( {\partial S_{0y} \over \partial y }\right)^{-2}
\left({\partial^2 S_{0y} \over \partial y^2 }\right)^2
 -\left( {\partial S_{0y} \over \partial y }\right)^{-1}
{\partial^3 S_{0y} \over \partial y^3 } \right] \hskip8mm \nonumber \\
+V_{y}(y) -E_y ={\hbar^2 c_2\over 8m}
\left( {\partial S_{0y} \over \partial y}\right)^{-2}\left[
c_2-4 {\partial^2 S_{0y} \over \partial y^2 }  \right] \ ,
\end{eqnarray}
\begin{eqnarray}
{1\over 2m} \left({\partial S_{0z}\over \partial z}\right)^2
-{\hbar^2\over 4m}
  \left[ {3 \over 2 } \left( {\partial S_{0z} \over \partial z }\right)^{-2}
\left({\partial^2 S_{0z} \over \partial z^2 }\right)^2
 -\left( {\partial S_{0z} \over \partial z }\right)^{-1}
{\partial^3 S_{0z} \over \partial z^3 } \right] \hskip8mm \nonumber \\
+V_{z}(z) -E_z ={\hbar^2 c_3\over 8m}
\left( {\partial S_{0z} \over \partial z }\right)^{-2}\left[
c_3-4 {\partial^2 S_{0z} \over \partial z^2 }  \right] \ .
\end{eqnarray}
In the left hand side of Eq. (49), we recognize the usual one-dimensional QHJE.
Since for an arbitrary potential $V_x(x)$,
$
c_1-4 {\partial^2 S_{0x} / \partial x^2 }  \neq 0,
$
the usual one-dimensional QHJE is a particular case of (49) which happens only if
$c_1=0$. Therefore, the decomposition (25) and (35) does not automatically
reproduce the usual one-dimensional case.
This means that the procedure consisting to search for the reduced action in the form
(25) as suggested in \cite{djama1} spreads confusion. This follows from a profound
geometrical origin. In fact, it is shown in \cite{BFM} that assuming the decomposition
(25) and (35) leads to express the quantum potential as a sum of Schwarzian
derivatives and does not provide a covariant formulation within the framework of
the equivalence postulate.

In conclusion, even with the assumption (25), the presence of the constants $c_1$, $c_2$
and $c_3$ in relations (49), (50) and (51) means that in the above expression (26)
of $S_0^{(2)}$ there is a lack of integration constants. So, in $S_0^{(2)}$ as suggested
in \cite{djama1}, there is a loss of information concerning the quantum motion.

We would like to add that the above solution (24) established in \cite{BM1} is really complete.
In fact, from the mathematical point of view, the SE is strictly equivalent to the couple of
equations (8) and (9) in the case where the wave function is complex. In this case, it is
established in \cite{BM1} that by fixing the initial
conditions for the most general solution of the SE, the above expression (24) contains
the exact number of pertinent parameters necessary to fix univocally the reduced action,
as it is well-known in one dimension. This represents a strong argument in favour of (24).
In the next section, we will come back to this problem of initial conditions.

\section {The quantum trajectories}

Trajectories in the context of quantum mechanics were introduced by Einstein \cite{holland, belousek},
de Broglie \cite{broglie} and Bohm \cite{bohm1,bohm2}. For various reasons described in the literature
\cite{floyd2,BD1,bouda2,Wyatt}, other formulations were published in the one-dimensional stationary case.
In \cite{BD1}, by appealing to the quantum transformation \cite{FM2,FM3} allowing to write the QHJE in
the classical form, the relation
\begin{equation}
{1 \over 2 }  {\partial S_{0} \over \partial x } \dot{x} + V(x) = E  \ ,
\end{equation}
is derived. This relation has allowed to establish the quantum Newton law. Although (52) works
in classical mechanics $(\partial S_{0}^{clas} / \partial x  = m \dot{x})$, it describes the quantum
motion because $\partial S_{0} / \partial x $ is the solution of the QHJE, Eq. (10).
The higher dimension version of (52) can be sensibly assumed as
\begin{equation}
{1 \over 2 }  {\partial S_{0} \over \partial x } \dot{x} +
{1 \over 2 }  {\partial S_{0} \over \partial y } \dot{y} +
{1 \over 2 }  {\partial S_{0} \over \partial z } \dot{z} +
                                                          V(x,y,z) = E  \ .
\end{equation}
What is at stake is how to prove this relation. In \cite{djama3}, the author claimed that he
provided a proof. However, his reasoning is false. In fact, it is obvious that the
identification of Eqs. (56) and (58) in \cite{djama3} is erroneous because the author compared
in a development with respect to derivatives of $S_0$ coefficients which are themselves
depending on these derivatives. In addition, the author used the wrong expressions of these
coefficients to determine trajectories in the three-dimensional constant potential case while
the above law (53) should not be enough to describe a motion in higher dimensions.
Even in classical mechanics, it is known that the use of the energy conservation equation is
insufficient to describe a motion except in one dimension.

We would like to indicate that a correct proof of (53) is provided in \cite{BG} in which
we stressed that this law must be completed in order to be applied in the case where the
dimension of space is more than one.

The last point that we would like to raise concerns microstates associated to various trajectories for
the same physical state. In any dimension, if the variables
are separable, the manifestation of microstates is occurred in the case where the system is
described by a real wave function, up to a constant phase factor. In the case of complex wave
function, it is established in \cite{BM1} that by fixing the initial conditions for the most
general solution of the three-dimensional SE,
\begin{equation}
\phi(x,y,z)=\sum_{i=1}^8 c_i \phi_i \ ,
\end{equation}
the reduced action is univocally fixed, $(c_i)$ being a set of eight complex constants and
$(\phi_i)$ the real solutions of the SE defined in (23). In others words, there is no
microstates. However, in the context of the Copenhagen interpretation and separated variables,
the probability density in the multidimensional configuration space is the product of individual
probability densities. This is the reason for which the wave function is often sought in the form
given by the above relation (16) in which $\phi_x(x)$, $\phi_y(y)$ and $\phi_z(z)$ are the
general solutions
\begin{equation}
\phi_x(x)= a_x X_1 + b_x X_2 \ , \ \ \ \ \
\phi_y(y)= a_y Y_1 + b_y Y_2 \ , \ \ \ \ \
\phi_z(z)= a_z Z_1 + b_z Z_2
\end{equation}
of the corresponding one-dimensional SEs. In (55),  $X_{1}$, $X_{2}$, $Y_{1}$, $Y_{2}$, $Z_{1}$
and $Z_{2}$ are the same functions as in (23) and $(a_x$, $b_x$, $a_y$, $b_y$, $a_z$, $b_z)$
is a set of complex constants. Substituting the one-dimensional general solutions (55) in (16),
we obtain the same form as in (54)
\begin{equation}
\phi(x,y,z)=\sum_{i=1}^8 c'_i \phi_i \ .
\end{equation}
Contrary to (54) in which the eight coefficients $c_i$ are independent, in (56) the $c'_i$ can
be expressed in terms of the six parameters $a_x$, $b_x$, $a_y$, $b_y$, $a_z$ and $b_z$. To be
more precise, a simple rearrangement shows that $c'_i$ can be expressed in terms of only four
independent parameters. Thus, fixing the initial conditions for the wave function and
following the same procedure as the one developed in \cite{BM1},
we cannot determine all the integration constants in (24) and then we do not fix completely
the reduced action. This reveals the presence of microstates. Contrary to the
one-dimensional case \cite{bouda1,floyd3}, we conclude that the QHJE describes microstates not
detected by the SE even in the case of a complex wave function if it is constructed from (16).

\section{ Conclusion}

To summarize, we pointed out in this paper some flaws in earlier publications
concerning in higher dimensions the construction of the reduced action as a solution of the QHJE
\cite{djama1,djama2} and the establishment of the quantum law of motion \cite{djama3}.
We also indicated how to correct these flaws and provided new insights into the investigation
of the QHJE. In particular, we proved that the reduced action constructed as a sum of one variable
functions, as in classical mechanics, does not contain a complete information about the quantum motion.
Finally, we established that in higher dimensions the QHJE describes microstates not detected
by the SE even in the case where the wave function is complex. From the mathematical point of view, we
stress to indicate that in the case where the wave function is complex, the QHJE is equivalent to the SE.
In fact, if the wave function is written as in (54) where the eight coefficients $c_i$ are
independent, there is no trace of microstates \cite{BM1}. However, the existence of various trajectories
for the same state described by a complex wave function is a consequence
of the relationships between the eight coefficients $c'_i$  appearing in (56) and imposed by
the Copenhagen interpretation.

\bigskip
\bigskip

\noindent
{\bf REFERENCES}


\begin{enumerate}

\bibitem{broglie}
L. de Broglie, An. Fond. Louis de Broglie, {\bf 12}, 1 (1987).

\bibitem{bohm1}
D. Bohm, Phys. Rev. {\bf 85},  166 (1952).

\bibitem{bohm2}
D. Bohm, Phys. Rev.  {\bf 85}, 180 (1952).

\bibitem{FM1a}
A. E. Faraggi and M. Matone, Phys. Lett. B {\bf 450}, 34 (1999), 
hep-th/9705108.

\bibitem{FM1b}
A. E. Faraggi and M. Matone, Phys. Lett. B  {\bf 437}, 369 (1998), 
hep-th/9711028.

\bibitem{FM2}
A. E. Faraggi and M. Matone, Int. J. Mod. Phys. A {\bf 15},  1869 (2000), 
hep-th/9809127.

\bibitem{BFM}
G. Bertoldi, A. E. Faraggi and M. Matone, Class. Quant. Grav. {\bf 17},
3965 (2000), hep-th/9909201.

\bibitem{poirier}
B. Poirier, J. Chem. Phys. {\bf 121}, 4501 (2004).

\bibitem{bouda1}
A. Bouda, Found. Phys. Lett. {\bf 14}, 17 (2001), quant-ph/0004044.

\bibitem{floyd1a}
E. R. Floyd, Phys. Rev. {\bf D34}, 3246 (1986).

\bibitem{floyd1b}
E. R. Floyd, Phys. Essays  {\bf 5}, 130 (1992).

\bibitem{BD1}
A. Bouda and T. Djama, Phys. Lett. A {\bf 285}, 27 (2001), quant-ph/0103071.

\bibitem{BM1}
A. Bouda and A. Mohamed Meziane, Int. J. Theo. Phys.  {\bf 45}, 1323 (2006), 
quant-ph/0701159.

\bibitem{djama1}
T. Djama, Phys. Scr. {\bf 75}, 77 (2007).

\bibitem{djama2}
T. Djama, arXiv e-print: quant-ph/0404175.

\bibitem{holland}
P. Holland, Found. Phys. {\bf 35}, 177 (2005), .

\bibitem{belousek}
D. W. Belousek, Stud. His. Phil. Mod. Phys. {\bf 27}, 437 (1996).

\bibitem{floyd2}
E. R. Floyd, Phys. Rev. D {\bf 26}, 1339 (1982).

\bibitem{bouda2}
A. Bouda, Int. J. Mod. Phys. A {\bf 18}, 3347 (2003), quant-ph/0210193.


\bibitem{Wyatt}
R. E. Wyatt, Quantum Dynamic with Trajectories,
Springer, ISBN: 0-387-22964-7 (2005).

\bibitem{djama3}
T. Djama, Phys. Scr. {\bf 76}, 82 (2007).

\bibitem{BG}
A. Bouda and A. Gharbi, Int. J. Theo. Phys.  {\bf 47}, 1068 (2008), 
arXiv:0810.0826.

\bibitem{FM3}
A. E. Faraggi and M. Matone, Phys. Lett. A {\bf 249}, 180 (1998), 
hep-th/9801033.

\bibitem{floyd3}
E. R. Floyd, Found. Phys. Lett. {\bf 9}, 489 (1996), quant-ph/9707051.

\end{enumerate}
\end {document}